\documentstyle[stwol]{article}

\def\ra{\rightarrow}

\def\tb{\tan \beta}

\def\be{\begin{equation}}
\def\ee{\end{equation}}
\def\bea{\begin{eqnarray}}
\def\eea{\end{eqnarray}}

\newcommand{\mr}{{\stackrel{<}{\sim}}}
\def\lsim{\:\raisebox{-0.5ex}{$\stackrel{\textstyle<}{\sim}$}\:}
\def\gsim{\:\raisebox{-0.5ex}{$\stackrel{\textstyle>}{\sim}$}\:}
\def\gev{\; \rm  GeV}
\def\eg{ {\it e.g. }}
\def\ie{ {\it i.e. }}
\def\bi{\bibitem}

\begin{document}

\title{\begin{flushright}
IFT-96/28\\ [1.5ex]
{\large \bf hep-ph/9612460 }\\ [1.5ex]
December 1996
\end{flushright}
~\
{STATUS OF 2HDM WITH A LIGHT HIGGS PARTICLE}
}

\author{ M. KRAWCZYK}

\address{Institute of Theoretical Physics, Warsaw University,
 Ho\.za 69,
Warsaw, 00-681, POLAND}



\twocolumn[\maketitle\abstracts{
Present data do not rule out the light neutral Higgs particle
$h$ or $A$ with mass below 40--50 GeV 
in the framework of the general 2HDM ("Model II"). 
The status of this model in a light of existing LEP I data
and a potential of the new muon 
experiment (g-2), 
 the measurement of   
photon-gluon and gluon-gluon fusion at HERA as well as  photon-photon 
fusion at
 low energy $\gamma \gamma$ NLC
is discussed.
}]

\footnote{To appear in the 
``Proceedings of the 28th International
 Conference on High Energy Physics'', Warsaw, Poland, 25-31 July 1996,
World Scientific, ed. Z.Ajduk, A. K. Wr\,oblewski}

\section{Introduction}
The mechanism of spontaneous symmetry breaking  proposed as
the source of mass for the gauge and fermion fields in the Standard 
Model (SM) leads to  a neutral scalar particle,
the minimal Higgs boson.  According to  the LEP I data,
based on the Bjorken process $e^+e^- \ra H Z^*$,
it   should be heavier than 66 GeV \cite{hi,lep},
also
the MSSM neutral Higgs particles have  been
constrained by LEP1 data to be heavier than 
  $\sim$ 45 GeV \cite{lep,susy,hi}. The general two Higgs doublet
model (2HDM) may yet accommodate a very light ($ \lsim 45-50 \gev$)
 neutral scalar $h$ {\underline {or}} a pseudoscalar $A$ as long as
$M_h+M_A \gsim M_Z$ ~\cite{lep} $^{, 3c)}$.
Note that the lower limit for the { {charged}} Higgs boson $M_{H^{\pm}}
$=
44 GeV/c was obtained at LEP I \cite{hi}
from process $Z \ra H^+H^-$ 
(moreover  in 
the MSSM version one expect 
$M_{H^{\pm}} > M_W$).

In the  minimal extension of the Standard Model there are 
two Higgs doublets,  
the observed Higgs sector is enlarged to five scalars: two
neutral Higgs scalars (with masses $M_H$ and $M_h$ for heavier and
lighter particle, respectively), one neutral pseudoscalar
($M_A$), and a pair of charged Higgses ($M_{H^+}$ and $M_{H^-}$). 
 The
neutral Higgs scalar couplings to quarks, charged leptons and gauge bosons
 are 
modified with respect to analogous couplings in SM by factors that 
depend on additional parameters : $\tan\beta$, which is
the ratio of the vacuum expectation values of the Higgs doublets
 $v_2/v_1$,
and the mixing angle in the neutral Higgs sector $\alpha$. Further,
 new couplings appear, e.g. $Zh (H) A$ and $ZH^+ H^-$ \cite{hunter}.
\subsection{The status of 2HDM Model after LEP I}\label{subsec:status}
In this talk I will  focus on the  "Model II" of the 
two Higgs doublet extentions of SM, where one Higgs doublet
with vacuum expectation value $v_2$ couples only to the "up"
components 
of fermion doublets while the other one couples to the "down" 
components  \footnote{In such model FCNC processes are absent 
and  the $\rho $ parameter retains its SM value at the tree level.
Note that in such scenario 
the large ratio $v_2/v_1 \sim m_{top}/m_b\gg 1$ is naturally 
expected. } \cite{hunter}. {{In particular,  fermions couple to 
the pseudoscalar $A$
with a strength  proportional to $(\tan \beta)^{\pm1}$
whereas the coupling of the fermions to the scalar $h$
goes as $\pm(\sin \alpha/\cos \beta)^{\pm1}$, where the sign
$\pm$  corresponds to  isospin $\mp$1/2 components}}. 
 
In the well known supersymmetric model (MSSM) belonging  to this  class
 the relations among the parameters required by the 
supersymmetry appear, leaving only two parameters free
(at the tree level) e.g. $M_A$ and $\tb$.
In general case, which we call the general 2 Higgs Doublet Model
 (2HDM), masses and parameters $\alpha$ and $\beta$ 
are not constrained by the model.
Therefore the same experimental data may lead to very different
consequences  depending on which 
 version of two Higgs doublet extension of SM,
supersymmetric or nonsupersymmetric, is considered (see below).
\begin{figure*}[ht]
\vskip 1.85in\relax\noindent\hskip .05in
	     \relax{\includegraphics{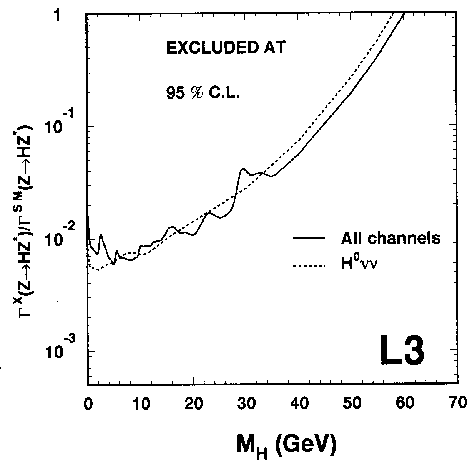}}
	     \relax\noindent\hskip 2.2in
	     \relax{\includegraphics{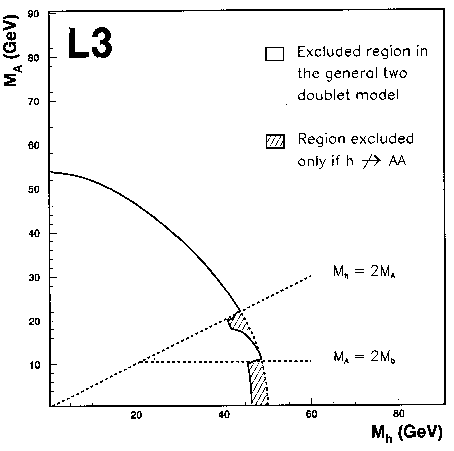}}
		     \relax\noindent\hskip 2.1in
	     \relax{\includegraphics{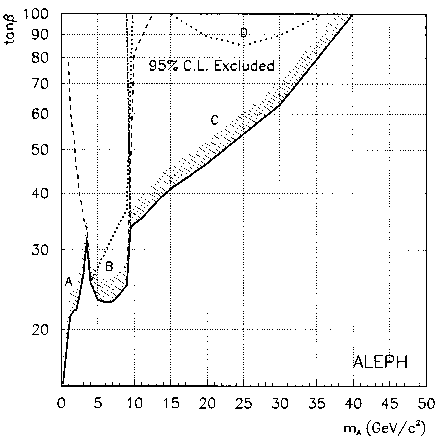}}
\vspace{-2.5ex}
\caption{ {\em a) The 95\% exclusion plot for   
scalar from the Bjorken process $^{2c)}$.
b)
The 95\% exclusion plot for  the mass of 
scalar versus  the mass of pseudoscalar  in 2HDM $^{3c)}$.
c) The 95\% exclusion plot for   
$\tan \beta$ versus the pseudoscalar mass from the Yukawa process.
Parameter space above the curves can be ruled out $^{7}$.
}}
 \label{fig:excl}
\end{figure*}

 For {\underline {neutral}} Higgs particles $h$ and 
$A$ there are two
	  main and complementary sources  of information at LEP I:
  the Bjorken processes $Z \ra Z^*h $,
which constrains  $g_{hZZ}^2 \sim \sin^2(\alpha-\beta)$
for $M_h$ below 50-60 GeV \cite{lep}(Fig. 1a)
and pair production   $Z\ra hA$,
constraining the $g_{ZhA}^2 \sim cos^2(\alpha-\beta)$ \cite{lep,susy}
for $M_h+M_A\mr M_Z$ 
{\footnote {
 The off shell production could also be included.}.
 The Higgs pair production cross section depends 
also on the masses $M_h$, $M_A$ and $M_Z$.
Combined results on $\sin^2(\alpha-\beta)$ and $\cos^2(\alpha-\beta)$   
can be translated into
the limits on neutral Higgs boson masses $M_h$ and $M_A$.
 In the MSSM, due to relations among parameters, 
the above data allow to draw limits for the masses
 of {\underline {individual}}
 particles: $M_h\gsim 45$ GeV for any $\tan \beta $ 
 and $M_A \gsim$ 45 GeV for $\tan\beta \ge$1 \cite{susy,hi}.
In the general 2HDM the implications are quite different, 
here the large portion of the ($M_h$,$M_A$) plane,
where {\underline {both}} masses are in the range between 
0 and $\sim$50 GeV, is excluded \cite{lep,susy} (Fig. 1b)
(for comparison   the corresponding  plot
 obtained for MSSM is presented in Fig.2).
\begin{figure}[hb]
\vskip 1.7in\relax\noindent\hskip .5in
	     \relax{\includegraphics{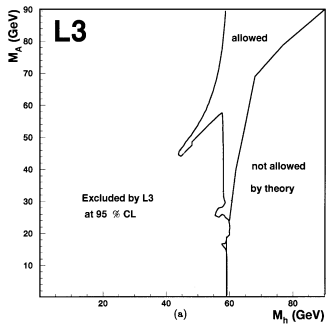}}
\vspace{-2.5ex}
\caption{ {\em  The allowed and excluded region of Higgs boson mass in  
the MSSM $^{3c)}$.  
}}
\label{fig:mass}
\end{figure}
The third basic  process  in  search of a neutral  
Higgs particle at LEP I is the Yukawa process, $\ie$
 the bremsstrahlung production   of the neutral Higgs
boson $h(A)$ from the heavy fermion, 
$e^+e^- \rightarrow f {\bar f} h(A)$, where $f$ means here
{\it b} quark or $\tau$
lepton.
This process plays a very important role since 
it  constrains the production of a very light pseudoscalar
even if the pair production is forbidden kinematically,
 $\ie$ for $M_h+M_A>M_Z$ {\footnote{for the off shell production}}.  
It allows also to look for a light scalar, being an additional,
and in case of $\alpha=\beta$ the most important, source of information
 \cite{pok,gle,kk}.    
{{ New analysis of the Yukawa process by  
 ALEPH collaboration, contributed to this conference \cite{alef}, 
 led to  the exclusion  plot (95\%)   on the $\tb$ versus the  
pseudoscalar mass, $M_A$ (Fig.1c).
 The obtained  limits are rather weak
 allowing for {\sl the existence of a light $A$ with large $\tb$
(for mass below 10 GeV
  $\tb $ till 20--30, whereas 
 for $M_A$=40 GeV $\tb$ up to  100 is allowed!) }
For scalar  $h$ similar exclusion limits should 
 hold also (with
the replacement in coupling $\tb\ra \sin\alpha/\cos\beta$)
\footnote {Larger differences  one should expect however 
in region of  mass below 10 GeV
where more 
stringent limits should be obtained \cite{kk}.}.

As far as  other experimental data, 
especially from  low energy measurements, 
they 
cover only part of the parameter space of 2HDM, moreover
some of them  like the Wilczek process
 have large theoretical uncertainties
due both to  the QCD and relativistic corrections \cite{wil,hunter,bk}.

In light of the above experimental 
results 
 there is still the possibility of the
 existence of one
light neutral Higgs particle with mass below $\sim$ 40--50 GeV in 2HDM.
In the following we will study this model assuming,
according to  LEP I data,
the following mass relation between the lightest
neutral Higgs particles: $M_h+M_A \ge M_Z$.
 We specify the model further  by choosing particular 
values for the parameters $\alpha$ and $\beta$
within the present limits from
LEP I, we simply  take  $\alpha=\beta$.

As we described above the existing 
limits for
a light  neutral Higgs scalar/pseudoscalar 
boson  in 2HDM are rather weak. 
Therefore it is extremely important to check if  more stringent
limits can be obtained from  other measurements \cite{gle,ames,g22}.
\subsection{Constraints on  the parameters of 2HDM from present
$(g-2)$ data for muon.}
The present experimental data limits on $(g-2)$ for muon, 
averaged over the sign of the muon electric charge,
 is given by \cite{pres,data}:
$$a_{\mu}^{exp}\equiv{{(g-2)_{\mu}}\over{2}}=1~165 ~923~(8.4)\cdot 
10^{-9}.$$
The quantity within parenthesis, $\sigma_{exp}$, refers to the uncertainty
in the last digit.
 The theoretical (SM) result
$$a_{\mu}^{SM}=a_{\mu}^{QED}+a_{\mu}^{had}+a_{\mu}^{EW},$$
has error, mainly due to  the hadronic contribution, which is smaller than
 $\sigma_{exp}$. Still there is a large discrepancy
between theoretical results.
We will consider here \cite{g22} so called case A, 
based on Refs.\cite{qed,ki,mar,mk,ll,czar},
with  relatively small error in the hadronic 
part and  case B (Refs.
\cite{ki,jeg,hayakawa,czar}) corresponds to  the two times  
 larger error in the hadronic part
  \cite{nath}:
$$
\begin{array}{lrr}
case  &~{\rm {A~[in}}~ 10^{-9}]&~{\rm {B~[in}}~ 10^{-9}] \\  
\hline
{\rm QED}   &~1 ~165~847.06 ~(0.02)
  &~1 ~165~847.06 ~(0.02) \\
{\rm had}   & 69.70 ~(0.76) &  68.82 ~(1.54)  \\
{\rm EW}    & 1.51 ~(0.04)  &  1.51 ~(0.04)   \\
\hline
{\rm tot}   &1~165~9 1 8.27 ~(0.76)   & 1 ~165~9 1 7.39 ~(1.54)
\end{array} 
$$
The room for new physics
\footnote{see discussion in Refs.\cite{gle,nath,ames,g22}}
 is given basically 
by  the difference between the experimental data and theoretical SM
prediction: $a_{\mu}^{exp}-a_{\mu}^{SM}\equiv \delta a_{\mu}$
{\footnote {However in the calculation of
  $a_{\mu}^{EW}$ the (SM) Higgs scalar 
contribution is included(see discussion in\cite{g22}).}}.
 Below the difference $\delta a_{\mu} $ 
 for these two cases, A and B,
is presented together with 
 the error $\sigma$, obtained by adding the experimental 
and theoretical errors in quadrature; also the 95\% C.L. limits  are shown:  
$$
\begin{array}{lcc}
case&{\rm{A~[in}}~10^{-9}]&{\rm{B~[in}}~10^{-9}]\\  
\hline
\!\delta a_{\mu}(\sigma)\!\!&\!\!4.73 (8.43)\!&\!5.61(8.54)\\
\hline
\!{lim}\!\!&\!\!-11.79\!\le\!\delta a_{\mu}\!\le\!21.25\!&\!-11.13
\!\le\!\delta a_{\mu}\!\le\!22.35\!\\
\!{lim_{\pm}}\!\!&\!\!-13.46\!\le\!\delta a_{\mu}\!\le\!19.94\!&
\!-13.71\!\le\!\delta a_{\mu}\!\le\!20.84
\end{array} 
$$
One can see that at 1 $\sigma$  level the difference $\delta a_{\mu}$
can be of positive and negative  sign.
For that  scenarios
 in which both positive and negative 
$\delta a_{\mu}$ may appear,
 the 95\% C.L. bound can be calculated in a straightforward way
(denoted above  by $lim$). 
For the model where  the contribution of
only {\underline {one}} sign 
is physically accessible ($\ie$ positive or negative $\delta a_{\mu}$),
 the other sign being unphysical, the 95\%C.L. limits
should be calculated  \cite{data}
 separately for positive and 
for  negative contributions ($lim_{\pm}$).

We will use above bounds for the constraining the 2HDM: 
 so we take  $\delta a_{\mu}= a_{\mu}^{(2HDM)}$ with  
contribution from   scalar $h$ ($a_{\mu}^h$), pseudoscalar $A$  ($a_{\mu}^A$)
 and  charged Higgs boson $H^{\pm}$ ($a_{\mu}^{\pm}$)
(``full'' 2HDM contribution,
relevant formulae \cite{haber}can be found in the Appendix in Ref.\cite{g22}).
Each of these terms
disappears in the limit of large mass, 
at small mass the contribution reaches its maximum (or minimum if negative)
value.
The scalar contribution $a_{\mu}^h(M_h)$ is positive whereas the
pseudoscalar  boson $a_{\mu}^A(M_A)$ gives
a negative contribution, also
the  charged Higgs boson contribution is 
negative. 

In the following
 we assume, according to the 
 LEP I, mass limits for charged nad neutral Higgs particles
in following way:         
for {\sl scenario  { a)}} with a light pseudoscalar  
we take  
$a_{\mu}^{(2HDM)}(M_A)= 
a_{\mu}^A(M_A)+ a_{\mu}^h(M_h^0\!=\!M_Z\!-\!M_A) 
+ a_{\mu}^{\pm}(M_{\pm}^0)$, while  
 for a light scalar {\sl {scenario  b)}}:
$a_{\mu}^{(2HDM)}(M_h)= 
a_{\mu}^h(M_h)+ a_{\mu}^A(M_A^0=M_Z-M_h) 
+ a_{\mu}^{\pm}(M_{\pm}^0)$. Here  mass limit of $H^{\pm}$
was used $M_{\pm}^0$=44 GeV.
Since  the contribution $  a_{\mu}^{(2HDM)}$
is for the  scenario {\sl b)} positive,
 whereas for the scenario {\sl a)}
negative
 we will use bounds provided by $lim_{\pm}$
introduced above. In the contrast to the ``full'' 2HDM contribution
the {\sl simple} approach is  based 
on only pseudoscalar or scalar contribution in case {\sl a)} or
case {\sl b)}, respectively.
It reproduces the full 2HDM prediction below mass say 30 GeV (see \cite{g22}).

The case A 
gives  more stringent $lim_{\pm}$
 for both positive and negative $\delta a_{\mu}$ (see table),
therefore this case was used in constraining parameters of the 2HDM.  
The obtained 95\%C.L. exclusion plots for $\tb$ for 
light $h$ or  $A$  
is presented in Fig. 3.
If one compare the upper curve resulting from the present  
$g-2$ data with the ALEPH results \cite{alef}  
one can find that in the pseudoscalar case
there appear additional restriction for  mass below 2 GeV.
 Still $\tb$ about 10-15 is allowed for mass around 1 GeV.
Case B will lead to similar limits with the rescaling curves
by factor 1.022(1.009) for a light scalar (a light pseudoscalar) case. 
\begin{figure}[h]
\vskip 2.5in\relax\noindent\hskip -1.7in
	     \relax{\includegraphics{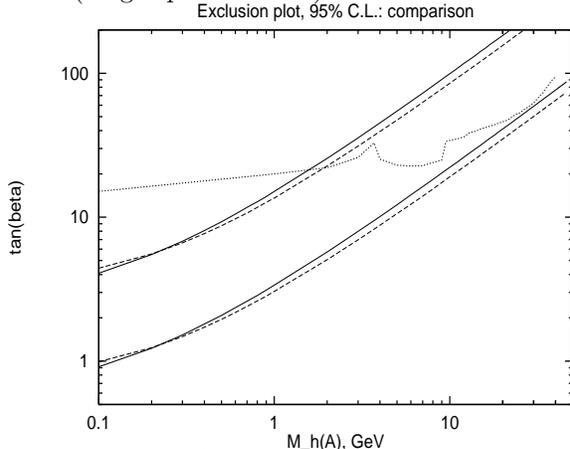}}
\vspace{-8.5ex}
\caption{ {\em The 95\% exclusion plots for  light 
scalar(solid lines) or light pseudoscalar (dashed lines)
in 2HDM (simple approach)$^{12}$. 
The limits {derivable} from present 
$(g - 2)_{\mu}$ measurement (upper lines) and from existing 
LEP I results (Yukawa process) for the pseudoscalar 
(dotted line) are shown. The lower lines correspond to the future 
(g-2) data (see text).
Parameter space above the curves can be ruled out.
}}
 \label{fig:excl}
\end{figure}
\section{ Potential of future experiments }
The role of future measurements:  (g-2) for muon,  
photon/gluon-gluon fusion at HERA collider and 
 $\gamma \gamma$ fusion at 
low energy NLC are discussed.
\subsection{ Future (g-2) data for muon}
Since presently  the dominant uncertainty in $\delta a_{\mu}$ 
is due to the experimental error,
the role of the forthcoming E821 experiment is  crucial
in testing the SM or  probing  a new physics.
 The expected new high-precision E821 Brookhaven
experiment has design sensitivity of $\sigma_{exp}^{new}=
4\cdot 10^{-10}$ (later even 1--2 $\cdot 10^{-10}$, see Ref.\cite{czar})
 instead of the above $84\cdot 10^{-10}$ (see talk at this conference
presented by B. L. Roberts\cite{fut}).
It is of great importance to reach similar accuracy in the theoretical
analysis.
One  expects  the improvement
in the calculation of the hadronic contribution
{\footnote {The improvement in the  ongoing experiments at low energy 
in expected as well.}} such
that the total uncertainty will be basically
due to the experimental error.
Below we will assume that 
the accessible range for the beyond SM contribution, 
in particular 2HDM  with  a light scalar or pseudoscalar,
would be smaller by factor 20 as compared with the present
$lim_{\pm}$95\% bounds for case A (Sec.1.2).
So, we consider the following option for future
(g-~2) measurement (in $10^{-9}$):
$${\rm lim_{\pm}}^{new}(95\%): -0.69\le\delta a_{\mu}^{new} \le 1.00.  $$
 The resulting exclusion plots for two scenarios in 2HDM
 obtained in the same manner
as in Sec. 1.2  ({\sl simple approach}) can be found in Fig. 3 
(lower curves).
They will be discussed together with others
exclusion plots in Sec. 2.3.
Here we would like only to mention that the assumed by us 
 $\delta a_{\mu}^{new}$ cover both positive and negative region,
but  
if the actual $\delta a_{\mu}^{new}$ will turn out to be 
 positive(negative) then the light
pseudoscalar(scalar) is no more allowed.
\subsection{ Photon-gluon and gluon-gluon fusion at HERA}
The gluon-gluon fusion
  via a quark loop,
$gg \ra h(A)$,
can be  a significant source of light non-minimal neutral Higgs bosons
at HERA  collider due to the hadronic
interaction of quasi-real photons with protons \cite{bk,hera}.
In addition the  production of the neutral Higgs boson via
$\gamma g \ra b {\bar b} h (A)$
may also be substantial \cite{grz,bk}. Note that the latter process
also includes  
 the lowest order contributions due to the resolved photon,
like $\gamma b\ra bh(A)$, $b {\bar b}\ra h(A)$, $bg \ra h(A)b$
etc. 
We study  the potential of both $gg$ and $\gamma g$ fusions
at HERA collider. 
It was found  that for mass below $\sim 30$  GeV the $gg$
fusion via a quark loop clearly dominates the cross section.
In order to detect the Higgs particle 
it is useful to
study the rapidity distribution ${d \sigma }/{dy}$ of the Higgs bosons
in the $\gamma p$ centre of mass system.
The (almost)
symmetric shape of  the rapidity distribution found for the signal
is extremely useful
 to reduce the
background and to separate the $gg\rightarrow h(A)$ 
contribution \cite{bk,hera}.

The main background for the Higgs mass range between
$\tau \tau$ and $b b$
 thresholds is due to $\gamma\gamma\rightarrow \tau^+\tau^-$.
In the region of negative rapidity 
the cross section ${d \sigma }/{dy}$ is very large, 
\eg ~for the $\gamma p$ energy equal to 170 GeV $\sim$ 800 pb
at the edge of phase space ($y\sim -4$), then it falls down rapidly
approaching $y=0$. At the same time signal reaches at most 10 pb
(for $M_h$=5 GeV).
The region of positive rapidity  is {\underline {not}} allowed 
kinematically for this process since here one photon interacts directly 
with 
$x_{\gamma}=1$, and therefore $y_{\tau^+ \tau^-}
 =-{{1}\over{2}}log{{1}\over{x_p}}\leq 0$.
 Moreover, there is a
 relation between rapidity and invariant mass:
 $M^2_{\tau^+ \tau^-}=e^{2y_{\tau^+ \tau^-}}S_{\gamma p}$.
Significantly different topology found for 
 $\gamma\gamma\rightarrow \tau^+\tau^-$ events
than for the signal  should
 allow to get rid of this background.
 The other sources of background are 
$q\bar q\rightarrow\tau^+\tau^-$ processes.
These processes contribute to positive and negative
rapidity $y_{\tau^+ \tau^-}$, with a flat and
relatively low cross sections in the central region (see for more details
Ref.\cite{bk,hera}).

To show the potential of HERA collider the exclusion plot
based on the $gg$ fusion via a quark loop 
can be obtained. In this case, as we mentioned above, 
it is easy to find the
part of the phase space where the background is negligible.
We focus on the ${\tau^+ \tau^-}$ final state
and to calculate the 95\% C.L. for allowed value of $\tb$
we take into account signal events corresponding only to 
the positive rapidity region (in the $\gamma p$ CM system).
Neglecting here the background   the number of events
were taken  equal to 3. 
 The  results
for the $ep$ luminosity ${\cal L}_{ep}$
=25 pb$^{-1}$ and 500 pb$^{-1}$
are presented in Fig. 4.
\subsection{Exclusion plots for 2HDM}
In Fig.4 the  95\% C.L. exclusion curves for the $\tb$ 
in the general  2HDM ("Model II") 
obtained by us  for a light scalar (solid lines)
and for a light pseudoscalar (dashed lines)
are presented in mass range below 40 GeV.
For comparison results from LEP I analysis (Yukawa process)
 for pseudoscalar is also shown (dotted line).
The region of ($\tb, M_{h(A)}$) above curves is excluded.
\begin{figure}
\center
\vskip 8.8cm
\relax\noindent\hskip -4.20in
\relax{\includegraphics{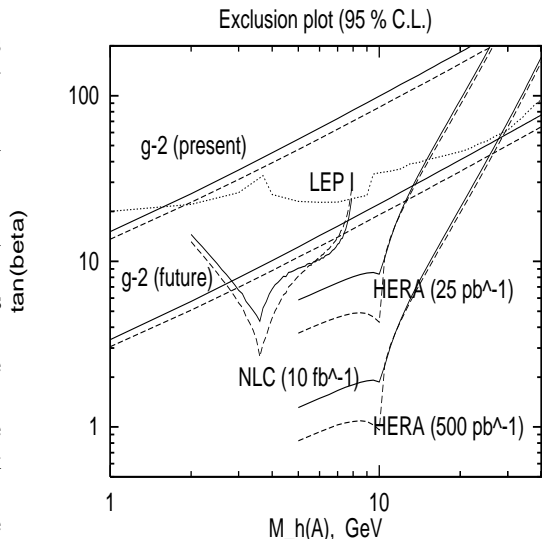}}    
\vspace{-12ex}
\caption{The combined exclusion plot.}
\label{fig:radk}
\end{figure}

 The present   $(g-2)_{\mu}$ data
 improve limits obtained recently by ALEPH collaboration 
 on $\tb$ for low mass of the pseudoscalar:  $M_A \lsim$ 2 GeV.
Similar situation  should  hold for a 2HDM with a light scalar,
although here the Yukawa process may be more restrictive for $M_h\lsim$
10 GeV \cite{kk}.
The  
future improvement      
in the accuracy  by factor 20 in the  forthcoming 
$(g-2)_{\mu}$ experiment  may lead to more stringent limits 
than provided by LEP I  up to mass of a neutral Higgs boson $h$ or $A$ 
equal to 30 GeV,
if the mass difference between scalar and pseudoscalar is $\sim M_Z$
(or to  higher $M_{h(A)}$ for a larger mass difference \cite{g22}).
Note however that there is some arbitrariness in the deriving the
expected bounds for the $\delta a_{\mu}^{new}$. 

The search at HERA in the gluon-gluon fusion via a quark loop
 at HERA may lead to even more stringent
limits  for the mass range 5--15 (5--25) GeV, provided
the luminosity will reach 25 (500) pb$^{-1}$ and the 
efficiency for $\tau^+ \tau^-$ final state will be high
enough \footnote{In this analysis the 100\% efficiency
has been assumed. If the efficiency will be 10 \% the corresponding 
limits will be larger by factor 3.3}.
The other production mechanisms like the $\gamma g$ fusion
and processes with the resolved photon are expected to improve
farther these limits. 

In the very low mass range the 
additional limits can be obtained from the low energy
 $\gamma \gamma$ NL collider. The results based 
on $\gamma \gamma \ra h(A)\ra \mu^+ \mu^-$
for the  luminosity ${\cal L}_{ee}$ 10 fb$^{-1}$ 
are presented in Fig.4 (from Ref.\cite{deb12}).
\section{Conclusion}
To conclude, in the framework of 2HDM 
a light neutral Higgs scalar or pseudoscalar,
in mass range below 40 GeV,
and $\tb$ even as large as  15-20 is not ruled out by the present data. 
The future experiments may clarify the status of the 
general 2HDM with the light neutral Higgs particle. 
The role of the forthcoming $(g-2)_{\mu}$ 
measurement seems to be crucial in
clarifying which scenario of 2HDM is allowed:  with a light scalar or
with a light pseudoscalar.
Then  farther constraints on the coupling of the allowed light Higgs
particle one can obtained from 
the HERA collider, which   is very well suitable  for this.
The very low energy region of mass may be studied
in addition in low energy $\gamma \gamma$ NLC machines. 
It is not clear however if the
low energy option will come into operation.
\section*{Acknowledgements}
I would like to thank to Jan \.Zochowski and Debayjoti Choudhury
for important contributions to this talk. 
Also I am indebted to Bernd Kniehl, 
Steve Ritz, JoAnna Hewett and Zygmunt Ajduk for their help.
Many thanks to 
J.-M. Grivaz, P. Janot and J.-B. de Vivie
for sending us the ALEPH results concerning the Yukawa process. 
Supported in part by the Polish Committee for Scientific Research.

\end{document}